\documentclass[11pt,twoside]{article}
\usepackage{asp2004}
\usepackage{psfig}
\usepackage{epsf}
\usepackage{graphics}
\usepackage{lscape}
\newcommand{\ie}{\emph{i.e.}\ }
\newcommand{\etal}{\emph{et al.}\ }
\newcommand{\eg}{\emph{e.g.}\ }

\newcommand{\kms}{km~s$^{-1}$}

\def\edcomment#1{\iffalse\marginpar{\raggedright\sl#1\/}\else\relax\fi}
\reversemarginpar

\begin{document}
\title{SN 1987A: The Unusual Explosion of a Normal  Type II Supernova}
\author{Nino Panagia}
\affil{ESA/Space Telescope Science Institute, 3700 San Martin Drive,
Baltimore, MD 21218, USA [panagia@stsci.edu]}

\begin{abstract}
I review the unwrapping story the SN~1987A explosion event, and
the  main discoveries associated with it. I will show that, although
this supernova is somewhat peculiar, the study of SN~1987A has
clarified quite a number of important aspects of the nature and the
properties of supernovae in general.
\end{abstract}
\thispagestyle{plain}
\section{Introduction}

Supernova 1987A was discovered on February 24, 1987 by Ian Shelton
(Kunkel \& Madore 1987) in the Large Magellanic Cloud.  SN~1987A is the
first supernova to reach naked eye visibility after the one studied by
Kepler in 1604 AD and is undoubtedly the supernova event best studied
ever by the astronomers. Actually, even if SN~1987A  has been more than
hundred times fainter than its illustrious predecessors in the last
millennium, it has been observed in such a detail and with such an
accuracy to make this event a {\it first} under many aspects (\eg
neutrino flux, progenitor identification, gamma ray flux) and definitely
the {\it best} studied of all. 

The early evolution of SN~1987A has been highly unusual and completely
at variance with the {\it wisest} expectations. It brightened much
faster than any other known supernova: in about one day it jumped from
12th up to 5th magnitude at optical wavelengths, corresponding to an
increase of about a factor of thousand in luminosity.  However, equally
soon its rise leveled off and took a much slower pace indicating that
this supernova would have never reached those high peaks in luminosity as
the astronomers were expecting.  Similarly, in the ultraviolet, the
flux initially was very high, even higher than in the optical.  But
since the very first observation, made with the International
Ultraviolet Explorer (IUE in short) satellite less than fourteen hours
after the discovery(Kirshner \etal 1987, Wamsteker \etal 1987), the
ultraviolet flux declined very quickly, by almost a factor of ten per
day for several days.  It looked as if it was going to be a quite
disappointing event and, for sure, quite peculiar, thus not suited to
provide any useful information  about {\it ``normal"} supernova
explosions.  But, fortunately, this proved not to be the case and soon
it became apparent that SN~1987A  has been the most valuable probe to
test our ideas about the explosion of supernovae.

Reviews of both early and recent observations and their
implications can be found in Arnett \etal (1989), McCray (1993, 2003,
2004),  Gilmozzi and Panagia (1999), and Panagia (2003). In the
following, I summarize some of the most  interesting findings 
from SN~1987A studies.

\begin{figure}[!ht]
\plotfiddle{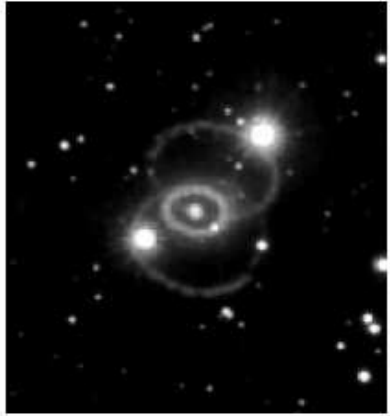}{6cm}{0}{30}{30}{-90}{-30}
\caption{True color picture ({\it HST-WFPC2}) of SN 1987A, its companion
stars, and the circumstellar rings
(Courtesy of Peter Challis)).}
\end{figure}

\section {Neutrino Emission from SN~1987A}
For the first time ever, particle emission from a supernova was
directly  measured from Earth: on February 23, around 7:36 Greenwich
time, the neutrino telescope "Kamiokande II" (a big cylindrical ``tub"
of water, 16~m in diameter and 17~m in height, containing about 3300
m$^3$ of water, located in the Kamioka mine in Japan, about 1000~m 
underground) recorded the arrival of 9 neutrinos within an interval of
2 seconds and 3 more 9 to 13 seconds after the first one.
Simultaneously, the same event was revealed by the IMB detector
(located in the Morton-Thiokol salt mine near Faiport, Ohio) and by
the  ``Baksan" neutrino telescope (located in the North Caucasus
Mountains, under Mount Andyrchi) which recorded 8 and 5 neutrinos, 
respectively,  within few seconds from each other.  This makes a total
of 25 neutrinos from an explosion that allegedly produces 10 billions
of billions of billions of billions of billions of billions of them!
But a little more than two dozens neutrinos was enough to
verify and confirm the theoretical predictions made for the core
collapse of a massive star (\eg Arnett \etal~ 1989 and references
therein). This process was believed to be the cause of the explosion of
massive stars at the end of their lives, and SN 1987A has provided the
experimental proof that the theoretical model was sound and correct,
promoting it from a nice theory to the description of the truth. 

\section {SN~1987A Progenitor Star} 

From both the presence of hydrogen in the ejected matter and the
conspicuous flux of neutrinos, it was clear that the star which had
exploded was quite massive, about twenty times more than our Sun. And
all of the disappointing peculiarities were due to the fact that just
before the explosion the supernova progenitor was a blue supergiant
star instead of being a red supergiant as common wisdom was predicting.
There is no doubt about this explanation because SN~1987A is exactly at
the same position as that of a well known blue supergiant,
Sk~$-69^{\circ}$~202. And the IUE observations indicated that such a
star was not shining any more after the explosion: the blue supergiant
star  unambiguously was the SN progenitor. This heretic possibility was
first suggested in Panagia \etal (1987) and subsequently confirmed
by the more detailed analyses presented by Gilmozzi \etal (1987) and
Sonneborn, Altner \& Kirshner (1987). 

On the other hand,  the presence of narrow emission lines of highly
ionized species, detected in SN~1987A short wavelength spectrum since late
May 1987,  has provided evidence for the progenitor having been a red
supergiant before coming back toward the blue side of the HR diagram
(Fransson \etal 1989). Also, the detection of early radio emission that
decayed in a few weeks (Turtle \etal 1987) indicated that the ejecta
were expanding within a circumstellar environment whose properties were
a perfect match to the expected wind of a blue supergiant progenitor
(Chevalier \& Dwarkadas 1995).

Such an evolution for an $\sim20~M_\odot$ star was not expected, and
theorists struggled quite a bit to find a plausible explanation for
it.  As summarized by Podsiadloswki (1992), in order to explain all
characteristics of SN~1987A, rotation  has to play a crucial role, thus
limiting the possibilities to models involving either a rapidly
rotating single star (Langer 1991), or a stellar merger in a massive 
binary system (Podsiadloswki 1992).

\begin{figure}[!ht]
\plotfiddle{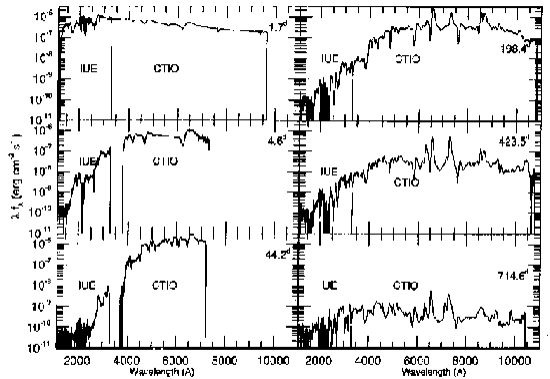}{5cm}{-90}{30}{30}{-135}{170}
\caption{Evolution of the UV and optical spectrum of SN~1987A 
(Pun et al. 1995).}
\end{figure}

\section{Explosive Nucleosynthesis} 

The optical flux reached a maximum around mid-May, 1987, and declined
at a quick pace until the end of June, 1987, when rather abruptly it
slowed down, settling on a much more gentle decline of about 1\% a day
(Pun  \etal~1995).  Such a behaviour was followed for about two years
quite regularly: a perfect constant decay with a characteristic time of
114 days, just the same as that of the radioactive isotope of cobalt,
$^{56}$Co, while transforming into iron. This is the best evidence for
the occurrence of nucleosynthesis during the very explosion: $^{56}$Co
is in fact the product of $^{56}$Ni decay and this latter can be formed at the
high temperatures which occur after the core collapse of a massive
star.  Thus, not only are we sure that such a process is operating in a
supernova explosion, but we can also determine the mass of Ni  produced
in the explosion, slightly less than 8/100 of a solar mass or $\sim$1\%
of the mass of the stellar core before the explosion.  The detection of 
hard X-ray
emission since July 1987, and the subsequent detection of
gamma-ray emission have confirmed the reality of such a process and
provided more detailed information on its distribution within the
ejecta (\eg Arnett \etal~1989 and references therein). Eventually, the
detection of Ni lines in the near infrared (Spyromilio \etal 1990)
confirmed the light curve result and provided the first {\it direct}
evidence of the production of  $^{56}$Ni in supernova explosions.

\section{Energetics of the Emitted Radiation} 
A catalog of SN~1987A ultraviolet spectra obtained with {\it IUE} (751
spectra over the period 1987 February 24 [day 1.6] through 1992 June 9
[day 1567]) have been presented by Pun \etal (1995). They show that the
UV flux plummeted during the earliest days of observations (Fig.~2)
because of the drop in the photospheric temperature and the increase in
opacity. However, after reaching a minimum of 0.04\% on day 44, the UV
flux increased by 175 times in its contribution to 7\% of the total
UVOIR bolometric luminosity at day 800 (Fig.~3). A study of the UV
colors reveals that the supernova started to get bluer in the UV around
the time when dust started to form in the ejecta. These results are
consistent with the possibility that the dust condensed may be
metal-rich.

\begin{figure}[!ht]
\plotfiddle{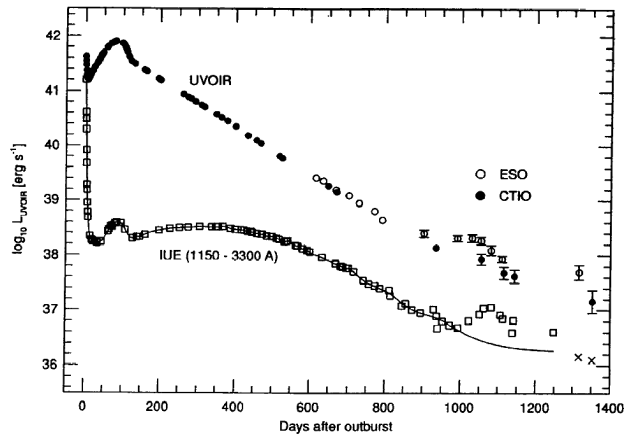}{5cm}{0}{30}{30}{-135}{-170}
\caption{ UV and bolometric light curve of SN~1987A 
(Pun et al. 1995 ).}
\end{figure}

\section{The UV Echo of SN~1987A: Spectrum of the Explosion} 
Bright transient events such as nova and supernova outbursts can give
rise to the phenomenon of a light echo. This is produced when light
from the explosion illuminates nearby interstellar dust and is
reflected in the direction of the observer. In the case of SN~1987A
echoes were predicted by Chevalier (1986) and Schaefer (1987), and
first detected by Crotts (1988) and Rosa (1988). Since the UV light
curve was already plummeting by the time of the first IUE observation, 
a UV echo is expected to be the reflection of the light emitted at the
very time of the explosion (i.e. {\it before} the discovery of the
supernova!).  Indeed, ultraviolet light emitted by SN~1987A at the
shock breakout was detected by means of  IUE  observations, made one
year apart from each other, at a location a few arcseconds outside a
bright portion of an optical echo (Gilmozzi \& Panagia 1999). The
spectrum of the echo shows a hot continuum and a wide P~Cyg-like
feature centered around 1500\AA\ (Fig.~4) which, if interpreted as
CIV~1550\AA, implies an expansion velocity at the time of the shock
breakout as high as 40,000 \kms. This is in agreement with the first ``direct"
IUE spectrum, taken 24 hours after the explosion, which showed a MgII
line with a terminal velocity of about 35,000 \kms (Kirshner \etal 1987). 

\begin{figure}[!ht]
\plotfiddle{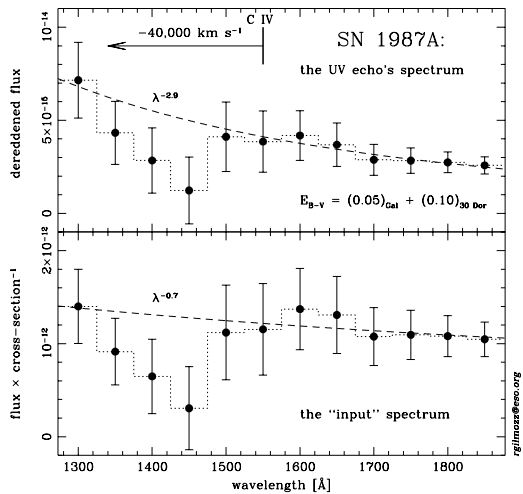}{6cm}{0}{35}{35}{-110}{-60}
\caption{The spectrum of the UV echo. {\bf Upper panel:} dereddened
observed data. {\bf Lower panel:} the effect of dust scattering 
 has been ``removed'' to reveal the actual shape of the spectrum
at the shock breakout. }
\end{figure}

\section{HST Observations - Structure and Expansion of the Ejecta} 

The Hubble Space Telescope ({\it HST}) was not in operation when the
supernova exploded, but it did not miss its opportunity in due time and
its first images, taken with the {\it ESA-FOC}~ on August 23 and 24, 1990,
revealed the inner circumstellar ring in all its ``glory" and detail
({\it cf.} Jakobsen \etal 1991), showing that, even with spherical
aberration,  {\it HST} was not a complete disaster, after all. More
observations were made with the {\it FOC}, which allowed Jakobsen \etal
(1993, 1994) to  measure the angular expansion of the supernova
ejecta.  The results confirmed the validity of the expansion models put
forward on the basis of spectroscopy. Additional observations, made
with the {\it WFPC2} on the re-refurbished {\it HST} confirmed the
early trend of the expansion and revealed the presence of structures
that had never been seen before (Jansen \& Jakobsen 2001, Wang \etal
2002). 

{\it HST-FOS} spectroscopic observations of SN~1987A, made over the
wavelength range 2000-8000 \AA~ on dates 1862 and 2210 days after the
supernova outburst, indicate that at late times the spectrum is formed
in a cold gas that is excited and ionized by energetic electrons from
the radioactive debris of the supernova explosion (Wang \etal 1996). The
profiles are all asymmetric, showing redshifted extended tails with
velocities up to 10,000 \kms~ in some strong lines. The blueshift of the
line peaks is attributed to dust that condensed from the SN~1987A ejecta
and is still distributed in dense opaque clumps.

\begin{figure}[!ht]
\plotfiddle{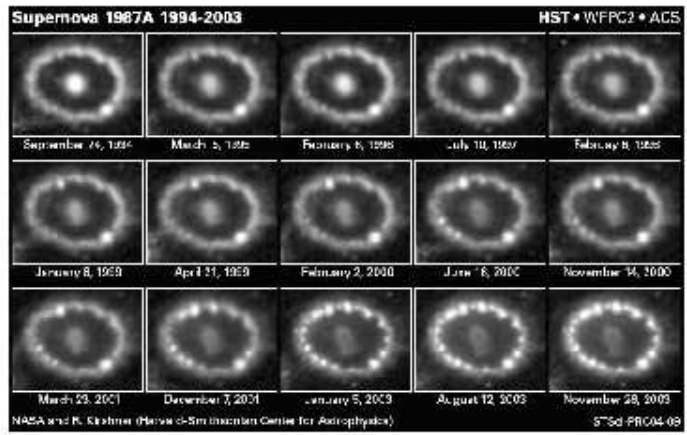}{7cm}{-90}{40}{40}{-160}{+230}
\caption{Series of images of SN 1987A and its inner circumstellar ring
obtained with {\it HST-WFPC2} between September 1994 and November 2003. 
It appears that the quiescent ring has developed at least twenty hot
spots in the last seven years. [Courtesy of  R.P Kirshner (Harvard) and
NASA]}
\end{figure}

\section{Properties and Nature of the Circumstellar Rings}   The study
of the circumstellar rings, \ie an equatorial ring (the ``inner ring")
about 0.86" in radius and inclined by about 45 degrees, plus two
additional "outer rings" which are approximately but not exactly,
symmetrically placed relative to the equatorial plane, approximately
co-axial with the inner ring, and have sizes 2-2.5 larger than the
inner ring. The presence of the inner ring was originally revealed with
the {\it IUE} detection of narrow emission lines (Fransson \etal 1989).
Heroic efforts done with ground based telescopes ( Crotts \etal 1989,
Wampler \etal 1990) provided early measurements of the shapes of the
circumstellar rings. Subsequently the rings were superbly imaged by
{\it HST} using both the {\it FOC} and {\it WFPC2} cameras (\eg 
{Jakobsen \etal 1991, Burrows \etal 1995).  Detailed studies of the
rings, mostly based on spectroscopy and imaging done with {\it HST},
have suggested that the rings, characterized by strong N
overabundances  (Fransson \etal 1989, Panagia \etal 1991, Panagia \etal
1996, Lundqvist \& Fransson 1996, Sonneborn \etal 1997), were ejected
in two main episodes  of paroxysmal mass loss which occurred
approximately 10,000 (the inner ring) and 20,000 years (the outer
rings) before the supernova explosion, respectively (Panagia \etal
1996, Maran \etal 2000).

\begin{figure}[!ht]
\plotfiddle{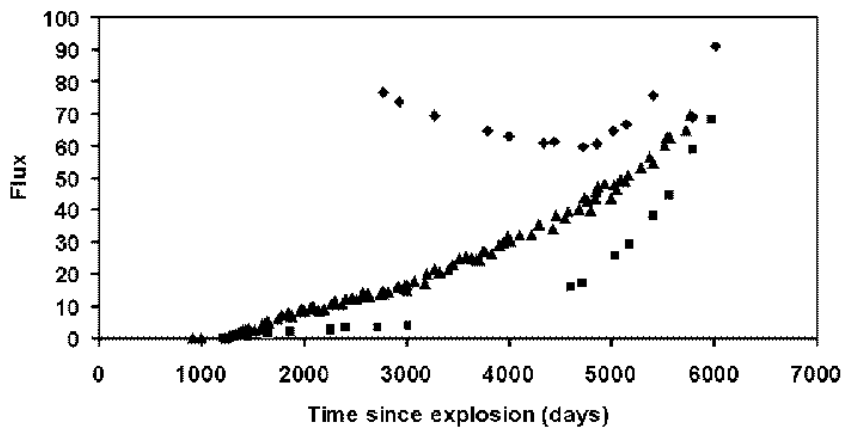}{6.6cm}{-90}{45}{45}{-185}{210}
\caption{The late evolution of the ring emission in the optical
(diamonds), radio (triangles) and X-ray (squares) domains [adapted from
McCray 2004]}
\end{figure}

\section{Interaction of the Ejecta with the Equatorial Ring}

Since mid-1997 Hubble has observed the high-velocity material from the
supernova explosion starting to overtake and crash into the slow-moving
inner ring. Figure~5 shows the dramatic evidence of these collisions.
The circumstellar ring started to develop bright spots in 1997, and  in
November 2003 one can identify at least twenty bright spots. These
bright spots are the result of the fast moving component of the ejecta
(at a speed of about  15,000 \kms) colliding with the stationary
equatorial ring (\eg Sonneborn \etal~1998, Michael \etal~2003).  
Independent evidence for an interaction whose strength is quickly
increasing with time is provided by both radio (Manchester \etal 2002)
and  X-ray (Park \etal 2002) emission ({\it cf.} Figure~6). Over the
next decades, as the bulk of the  ejecta  reach the ring, more spots
will light up and the whole ring will shine as it did in the first
several months after explosion (\eg McCray 2004).  Eventually, the
ejecta will completely sweep the ring up, clearing the circumstellar
space of that beautiful remnant of the pre-supernova wind activity. 
\section {SN~1987A:  An Ongoing Experiment }

It is clear that SN~1987A constitutes an ideal laboratory for the study
of supernovae, and of explosive events, in general. As summarized
above, a great deal of observations have been made and quite a number
of aspects have been clarified and understood.  At the same time, there
are still important points that need clarification and  further study,
as well as more observations. For example, the stellar remnant left
behind by the explosion has eluded our detection so far and its nature
remains a complete mystery.  Also, the detection of an early
interaction of the supernova ejecta with the inner circumstellar ring
has opened a new chapter in the study of this supernova, that is
expected to culminate in about ten years, when the colliding materials
will become the brightest objects in the LMC, with a display of
fireworks at X-ray, UV and optical wavelengths that defy our most vivid
imagination.


\begin{thebibliography}{}

\bibitem{arnetal89} Arnett, W.D.,  \etal 1989, ARA\&A, 27, 629

\bibitem{buetal95}Burrows, C.J., \etal 1995, ApJ, 452, 680

\bibitem{che86}Chevalier, R.A. 1986, ApJ, 308, 225

\bibitem{cd95}Chevalier, R.A., and Dwarkadas, V.V. 1995, ApJ, 452, L45

\bibitem {cro88}  Crotts, A. 1988, ApJ, 333, L51

\bibitem {ckmcc89cro88} Crotts, A., Kunkel, W.E., and McCarthy, P.J.
1989, ApJ, 347, L61

\bibitem{fretal}Fransson,  \etal 1989, ApJ, 336, 429

\bibitem{gil}Gilmozzi, R.,  \etal 1987, Nature, 328, 318

\bibitem{gipa}Gilmozzi, R., and Panagia, N., 1999, Mem.S.A.It., 70, 583

\bibitem{jak}Jakobsen, P., \etal 1991, ApJ, 369, L63

\bibitem {jaetal93}Jakobsen,  P., Macchetto, F.D., and Panagia, N. 1993
ApJ, 403, 736

\bibitem {jaetal94}Jakobsen,  P., Jedrzejewski, R.,  Macchetto, F.D.,
and Panagia, N. 1994, ApJ, 435, L47

\bibitem {jAJa01}Jansen, R.A., and Jakobsen, P. 2001, A\&A, 370, 1056

\bibitem{kir}Kirshner, R.P., \etal 1987, ApJ, 320,602

\bibitem{she87}Kunkel, W., and Madore, B. 1987, IAUC 4316

\bibitem {lf96}Lundqvist, P., and Fransson, C. 1996, ApJ, 464, 924

\bibitem{manetal02} Manchester, R.N., \etal 2002, PASAu, 19, 207

\bibitem{maretal00} Maran, S.P., \etal 2000, ApJ, 545, 390

\bibitem{mccra93}McCray, R. 1993,  ARA\&A, 31, 175

\bibitem{mccra03}McCray, R.  2003, in Supernovae and Gamma-Ray
Bursters,  ed. K. W. Weiler (Berlin: Springer-Verlag), p. 219-240

\bibitem{mccra04}McCray, R.  2004, in IAU Colloquium \#192
Supernovae, eds. J.M. Marcaide \& K.W. Weiler, (Berlin: Springer-Verlag), in
press

\bibitem{mietal03}Michael, E., \etal 2003, ApJ, 593, 809

\bibitem{pan87}Panagia, N., \etal 1987, A\&A, 177, L25 

\bibitem{pan91}Panagia, N., \etal 1991, ApJ, 380, L23

\bibitem{pan96}Panagia, N., \etal 1996, ApJ, 457, 604

\bibitem{panetal03}Panagia, N. 2003, in Supernovae and Gamma-Ray
Bursters,  ed. K.W. Weiler (Berlin: Springer-Verlag), p. 113-144

\bibitem{paretal02}Park, S., \etal 2002, ApJ, 567, 314

\bibitem{pun95}Pun, C.S.J., \etal 1995, ApJS, 99, 223

\bibitem {punetal02}Pun, C.S.J., \etal 2002, ApJ, 572, 906

\bibitem{ros88} Rosa, M. 1988, IAUC 4564

\bibitem{scha87} Schaefer, B.E. 1987, ApJ, 323, L47

\bibitem{sak87}Sonneborn, G., Altner, B., and Kirshner, R.P. 1987, ApJ,
323, L35
	
\bibitem{sonetal97} Sonneborn, G., \etal 1997, ApJ, 477, 848

\bibitem{sonetal98}Sonneborn, G., \etal 1998, ApJ, 492, 139

\bibitem{spyetal90}Spyromilio, J., \etal 1990, MNRAS, 242, 669

\bibitem{turetal87}Turtle, A.J., \etal 1987, Nature, 327, 38

\bibitem{wetal90} Wampler, J.E., \etal 1990, ApJ, 362, L13

\bibitem{wetal87} Wamsteker, W., \etal 1987, A\&A, 177, L21

\bibitem {wetal96}Wang, L., \etal 1996, ApJ, 466, 998

\bibitem {wetal02}Wang, L., \etal 2002, ApJ, 579, 671

\end{thebibliography}
\end{document}